\documentstyle[NATO]{crckapb} 

\input{psfig}

\def\Mpc{\, h^{-1} \, {\rm Mpc}}
\def\bfw{ {\bf w}}

\begin{opening}
\title{OBSERVATIONAL TESTS FOR THE COSMOLOGICAL PRINCIPLE AND WORLD MODELS}

\author{Ofer Lahav}
\institute{Institute of Astronomy\\
           Madingley Road, Cambridge CB3 0HA, UK}

\end{opening}

\runningtitle{THE COSMOLOGICAL PRINCIPLE}

\begin{document}


\begin{abstract}
  We review observational tests for the homogeneity of the Universe on 
 large scales. Redshift and peculiar velocity 
 surveys, radio sources, the X-Ray Background, the Lyman-$\alpha$ 
 forest
 and the Cosmic Microwave Background
 are used to set constraints on inhomogeneous models
 and in particular on fractal-like models.
 Assuming the Cosmological Principle
 and the FRW metric, we estimate cosmological parameters by joint analysis
 of  peculiar velocities, the CMB, cluster abundance, IRAS and  Supernovae.
 Under certain assumptions the best fit density parameter
 is $\Omega_{m} = 1 - \lambda \approx 0.4 $.
 We present a new method for joint estimation by combining different data sets
 in a Bayesian way, and utilising `Hyper-Parameters'.  
\footnote{
Review talk, to appear in the proceedings of the 
NATO ASI, Isaac Newton Institute, Cambridge, 
July 1999, ed. R. Crritenden \& N, Turok
Kluwer
} 
\end{abstract}

\section{Introduction}

The Cosmological Principle was first adopted when observational
cosmology was in its infancy; it was then little more than a
conjecture, embodying 'Occam's razor' for the simplest possible
model. Observations could not then probe to significant redshifts, the
`dark matter' problem was not well-established and the Cosmic Microwave
Background (CMB) and the X-Ray Background (XRB)  were still unknown.  
If the  Cosmological Principle turned out to be invalid 
then the consequences to our understanding of cosmology would be dramatic, 
for example the conventional way of interpreting the age of the Universe, 
its geometry and matter content would have to be revised. 
Therefore it is 
important to revisit this underlying assumption in the light of new
galaxy surveys and measurements of the background radiations.

Like with any other idea about the physical world, we cannot 
prove a model, but only falsify it.
Proving  the homogeneity of the Universe is in particular difficult 
as we observe the Universe from one point in space, and we can only 
deduce directly isotropy.
The practical methodology we adopt is to assume homogeneity and to assess
the level of fluctuations relative to the mean, and hence to test
for consistency with  the underlying hypothesis.
If the assumption of homogeneity turns out to be wrong, then 
there are  numerous possibilities
 for inhomogeneous models, and each of them must be
tested against the observations.

Despite the rapid progress in estimating the 
density fluctuations as a function of scale,
two gaps remain:

(i) It is still unclear how to relate the distributions of 
    galaxies and mass (i.e. `biasing');  
(ii) Relatively little is known about fluctuations 
on  intermediate scales 
between these of local galaxy surveys ($\sim 100 h^{-1} $ Mpc)
and the scales probed by COBE ($\sim 1000 h^{-1} $ Mpc).

Here we examine the degree of smoothness with scale 
by considering 
redshift and peculiar velocities surveys, 
radio-sources, the XRB, the Ly-$\alpha$ forest,
and the CMB.
We discuss some  inhomogeneous models
and show that a fractal model on large scales is highly improbable.
Assuming an FRW metric we evaluate the `best fit Universe' by 
performing a joint 
analysis of cosmic probes.

\section{Cosmological Principle(s)}

Cosmological Principles were stated over different periods in 
human history based on philosophical and aesthetic considerations 
rather than on fundamental physical laws.
Rudnicki (1995) summarized some of these principles in modern-day
language:

$\bullet$ The Ancient Indian: 
{\it The Universe is infinite in space and time and is 
infinitely heterogeneous}.

$\bullet$ The Ancient Greek:
{\it Our Earth is the natural centre of the Universe}.

$\bullet$ The Copernican CP:
{\it 
The Universe as 
observed from any planet looks much the same}.

$\bullet$ The Generalized CP:
{\it 
The Universe is (roughly) homogeneous and isotropic}.

$\bullet$ The Perfect CP:
{\it The Universe is (roughly) homogeneous in space and time,
and is isotropic in space}.

$\bullet$ The Anthropic Principle:
{\it A human being, as he/she is, can exist only in the Universe
as it is.}

\bigskip

We note that the Ancient Indian principle can be viewed as 
a `fractal model'. 
The Perfect CP led to the steady state model, 
which although more symmetric than the CP, 
was rejected on observational grounds.
The Anthropic Principle is  becoming popular again, e.g. in 
explaining a non-zero cosmological constant.
Our goal here is to quantify `roughly' in the definition of the 
generalized CP, and to assess if one may assume safely
 the Friedmann-Robertson-Walker
(FRW) metric of space-time.

\section {Probes of Smoothness}

\subsection {The CMB}  

The CMB is the strongest evidence for homogeneity.
Ehlers, Garen and Sachs (1968) showed that by combining the 
CMB isotropy with the Copernican principle 
one can deduce homogeneity. More formally the 
EGS theorem (based on Liouville theorem) states that
``If the fundamental observers in a dust spacetime see an isotropic
radiation field, then the spacetime is locally FRW''.
The COBE measurements of temperature 
fluctuations  $\Delta T/T = 10^{-5} $ on scales of $10^\circ$ give 
via the Sachs Wolfe effect ($\Delta T/T = \frac {1}{3} \Delta \phi/c^2$) 
and Poisson equation
rms density fluctuations of ${{\delta \rho} \over {\rho}} \sim 10^{-4} $ on $1000 \Mpc$ (e.g. Wu,  Lahav \& Rees 
1999; see Fig 3 here), i.e. the deviations from a smooth Universe are tiny.

\subsection {Galaxy Redshift Surveys}

 Figure 1 shows the distribution of galaxies in the ORS and IRAS 
 redshift surveys. It is apparent that the distribution is highly 
 clumpy, with the Supergalactic Plane seen in full glory.
 However, deeper surveys such as LCRS show that the fluctuations 
 decline as the length-scales increase. Peebles (1993) has shown 
 that the angular correlation functions for the Lick and APM surveys 
 scale with magnitude as expected in a universe which approaches 
 homogeneity on large scales.

\begin{figure}
\protect\centerline{
\psfig{figure=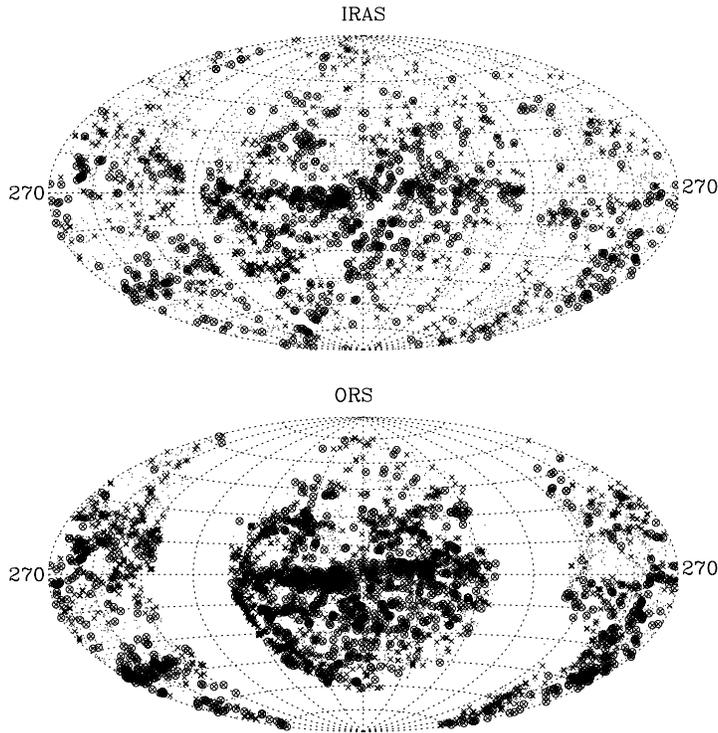,height=4truein,width=4truein}}
\caption[]{
The distribution of galaxies projected on the sky in
the IRAS and ORS samples.  This is an Aitoff projection in
Supergalactic coordinates, with $L = 90^\circ, B = 0^\circ$ (close to
the Virgo cluster) in the centre of the map. Galaxies within 2000 km/sec
are shown as circled crosses; galaxies between 2000 and 4000 km/sec are
indicated as crosses, and dots mark the positions of more distant
objects.  Here we include only catalogued galaxies, which is why the
Zone of Avoidance is so prominent in these two figures.
(Plot by M. Strauss, from Lahav et al. 1999).  }
\end{figure}

Existing optical and IRAS (PSCz) redshift surveys contain $\sim 10^4$
galaxies.  Multifibre technology now allows us to measure  redshifts
of millions of galaxies.  Two major surveys are underway.
The US Sloan Digital Sky Survey (SDSS) will measure redshifts to about
1 million galaxies over a quarter of the sky.  The Anglo-Australian 2
degree Field (2dF) survey will measure redshifts for 250,000 galaxies
selected from the APM catalogue.  About 80,000 2dF redshifts have been
measured so far (as of December 1999).  The median redshift of both the
SDSS and 2dF galaxy redshift surveys is ${\bar z} \sim 0.1$.  While
they can provide interesting estimates of the fluctuations on scales
of hundreds of Mpc's, the problems of biasing, evolution and
$K$-correction, would limit the ability of SDSS and 2dF to `prove' the
Cosmological Principle.  (cf. the analysis of the ESO slice by
Scaramella et al 1998 and Joyce et al. 1999).


\subsection{Peculiar Velocities}

Peculiar velocities are powerful as they probe directly the mass distribution
(e.g. Dekel et al. 1999).
Unfortunately, as distance measurements increase with distance, 
the scales probed are smaller than the interesting 
scale of transition to homogeneity.
Conflicting results 
on both the amplitude  and coherence of the flow suggest that
peculiar velocities cannot yet set strong constraints on the amplitude
of fluctuations on scales of hundreds of Mpc's.
Perhaps the most promising method for the future is the 
kinematic Sunyaev-Zeldovich effect which allows one to measure the 
peculiar velocities of clusters out to high redshift.

The agreement between the CMB dipole and the dipole anisotropy 
of relatively nearby galaxies argues in favour of large scale 
homogeneity.
The IRAS dipole (Strauss et al 1992, Webster et al 1998, 
Schmoldt et al 1999)
shows an apparent convergence 
of the dipole, with 
misalignment angle of only $15^\circ$.
Schmoldt et al. (1999) claim that 
2/3  of the dipole arises from within a $40 \Mpc$, 
but again it is difficult
to `prove' convergence from catalogues of finite depth.

 \subsection{Radio Sources}

Radio sources in surveys have typical median redshift
${\bar z} \sim 1$, and hence are useful probes of clustering at high
redshift. 
Unfortunately, it is difficult to obtain distance information from
these surveys: the radio luminosity function is very broad, and it is
difficult to measure optical redshifts of distant radio sources.
Earlier studies
claimed that  the distribution of radio sources supports the 
`Cosmological Principle'.
However, 
the wide range in intrinsic luminosities of radio sources
would dilute any clustering when projected on the sky (see Figure 2).  
Recent analyses  of
new deep radio surveys (e.g. FIRST)
suggest that radio sources are actually  clustered at least as strongly
as local optical
galaxies 
(e.g. Cress et al. 1996; Magliocchetti et al. 1998).
Nevertheless, on the very large scales the distribution of radio sources
seems nearly isotropic. 
Comparison of the measured quadrupole in a radio sample 
in the Green Bank and Parkes-MIT-NRAO
4.85 GHz surveys 
to the theoretically predicted ones (Baleisis et al. 1998)
offers a crude estimate of the fluctuations on scales $ \lambda \sim 600
h^{-1}$ Mpc.  The derived amplitudes are shown in Figure 3 for the two 
assumed Cold Dark Matter (CDM) models.
Given the problems of catalogue matching and shot-noise, these points should be
interpreted at best as `upper limits', not as detections.  

\begin{figure}
\protect\centerline{
\psfig{figure=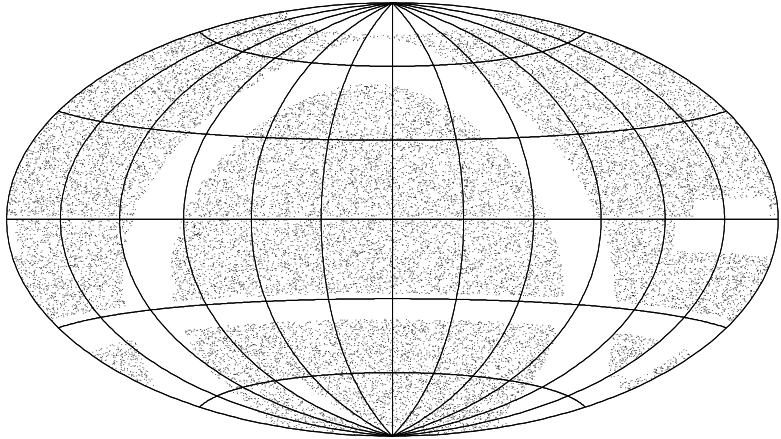,height=3truein,width=5truein}}
\caption[]{
The distribution of radio source from the 87GB and PMN surveys 
projected on the sky. 
This is an Aitoff projection in Equatorial coordinates 
(from Baleisis et al. 1998). }  
\end{figure}

\begin{figure}
\protect\centerline{
\psfig{figure=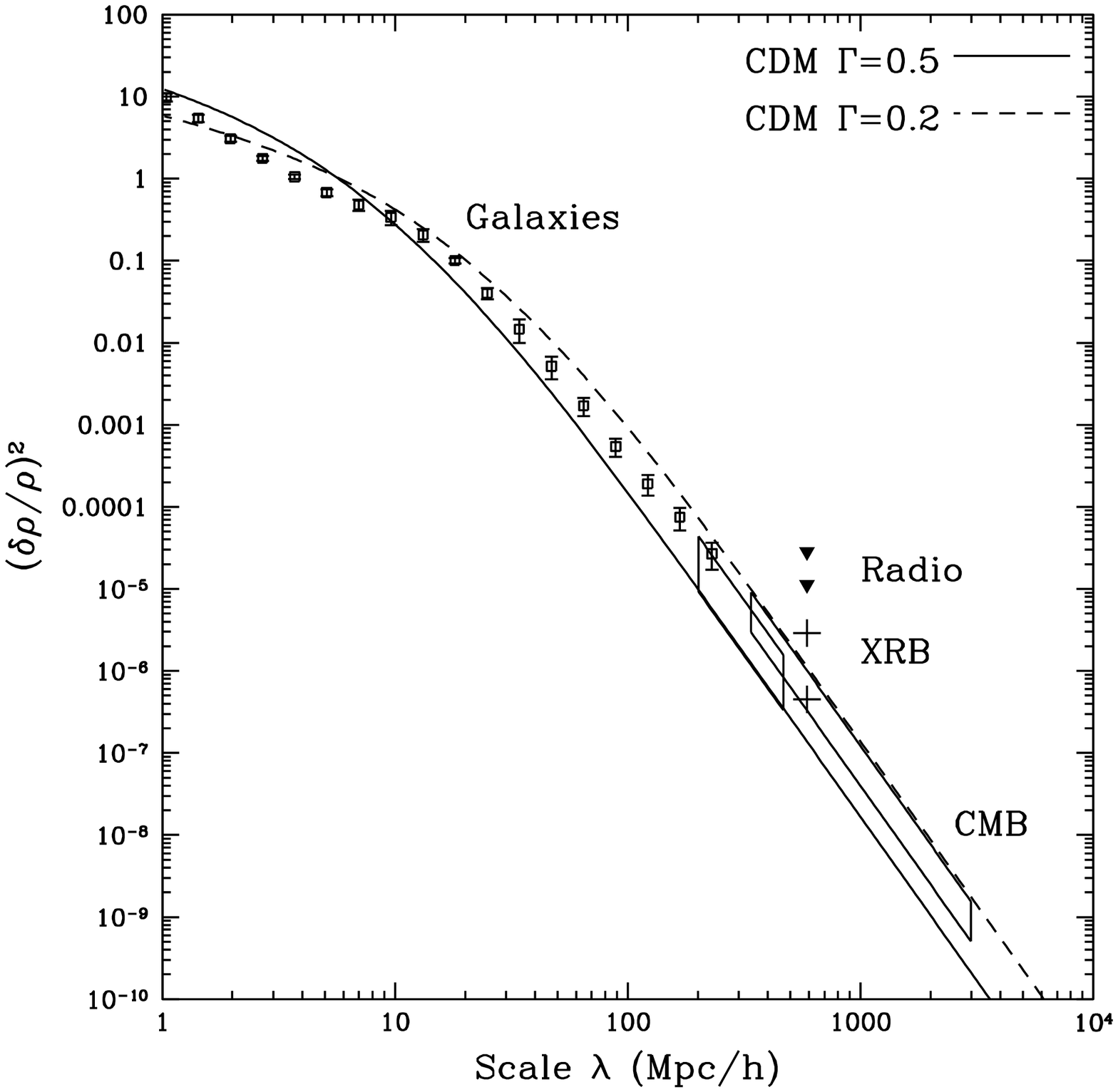,height=4truein,width=4truein}}
\caption[]{
  A compilation of density fluctuations on different scales from
  various observations: a galaxy survey, deep radio surveys, the X-ray
  Background and Cosmic Microwave Background experiments. The
  measurements are compared with two popular Cold Dark Matter models
  (with normalization $\sigma_8=1$ and 
  shape parameters $\Gamma=0.2$ and $0.5$). 
  The Figure shows mean-square density fluctuations $({ {\delta \rho}
    \over \rho })^2 \propto k^3 P(k)$, where $k=1/\lambda$ is the
  wavenumber and $P(k)$ is the power-spectrum of fluctuations.  
  The open squares at
  small scales are estimates from 
  the APM galaxy catalogue (Baugh \& Efstathiou 1994).
  The elongated 'boxes' at large scales represent the COBE
  4-yr (on the right) and
  Tenerife (on the left) CMB measurements  (Gawiser \& Silk
  1998).  The solid triangles and crosses represent amplitudes derived
  from the
  quadrupole of radio sources 
  (Baleisis et al. 1998) and the quadrupole of the XRB 
  (Lahav et al. 1997; Treyer et al. 1998).  
  Each pair of estimates corresponds to assumed shape of the two CDM models.
  (A compilation from Wu, Lahav \& Rees 1999).  }
\end{figure}

 \subsection {The XRB}

 Although discovered in 1962, the origin of
 the X-ray Background (XRB) is still unknown,  
 but is likely
 to be due to sources at high redshift 
 (for review see Boldt 1987; Fabian \& Barcons 1992).
 The XRB sources are probably
 located at redshift $z < 5$, making them convenient tracers of the mass
 distribution on scales intermediate between those in the CMB as probed
 by COBE, and those probed by optical and IRAS redshift
 surveys (see Figure 3).

The interpretation of the results depends somewhat on the nature of
the X-ray sources and their evolution.  
By comparing
the predicted multipoles to those observed by HEAO1 
(Lahav et al. 1997; Treyer et al. 1998; Scharf et al. 1999)
we estimate the amplitude of fluctuations for an
assumed shape of the density fluctuations 
(e.g. CDM models).  
Figure 3 shows the amplitude of fluctuations derived at the 
effective scale $\lambda \sim 600 h^{-1}$ Mpc probed by the XRB. 
The observed fluctuations in the XRB
are roughly as expected from interpolating between the
local galaxy surveys and the COBE CMB experiment.
The rms fluctuations 
${ {\delta \rho} \over {\rho} }$
on a scale of $\sim 600 h^{-1}$Mpc 
are less than 0.2 \%.

\subsection {The Lyman-$\alpha$ Forest}


The Lyman-$\alpha$ 
forest reflects the neutral hydrogen distribution and therefore
is likely to be a more direct  trace of the mass distribution 
than galaxies are.
Unlike galaxy surveys which are
limited to the low redshift Universe, the forest spans a large
redshift interval, typically $1.8 < z < 4$, corresponding 
to comoving interval of $\sim 600 \Mpc$.
Also, observations of the
forest are not contaminated by complex selection effects such as those
inherent in galaxy surveys.  It has been suggested qualitatively by
Davis (1997) that the absence of big voids in the distribution of 
Lyman-$\alpha$
absorbers is inconsistent with the fractal model.
Furthermore, all lines-of-sight towards quasars look
statistically similar.  
Nusser \& Lahav (1999) 
predicted the distribution of the flux  in Lyman-$\alpha$ 
 observations in a specific
truncated fractal-like model. They found that indeed in this model there are
too many voids compared with the observations and conventional (CDM-like)
models
for structure formation.
This too supports the common view that on large scales the Universe 
is homogeneous.

\section {Is the Universe Fractal ?}

The question of whether the Universe 
is isotropic and homogeneous on large scales
can also be  phrased in terms of the fractal structure of the 
Universe.
A fractal is a geometric shape that is not homogeneous, 
yet preserves the property that each part is a reduced-scale
version of the whole.
If the matter in the Universe were actually 
distributed like a pure fractal on all scales then the 
Cosmological Principle 
would be invalid, and the standard model in trouble.
As shown in Figure 3
current data already strongly constrain any non-uniformities in the 
galaxy distribution (as well as the overall mass distribution) 
on scales $> 300 \Mpc$.

If we count, for each galaxy,
the number of galaxies within a distance $R$ from it, and call the
average number obtained $N(<R)$, then the distribution is said to be a
fractal of correlation dimension $D_2$ 
if $N(<R)\propto R^{D_2}$. Of course $D_2$
may be 3, in which case the distribution is homogeneous rather than
fractal.  In the pure fractal model this power law holds for all
scales of $R$.

The fractal proponents (Pietronero et al. 1997)  have
estimated $D_2\approx 2$ for all scales up to $\sim 500\Mpc$, whereas
other
groups 
have obtained scale-dependent values 
(for review see Wu et al. 1999 and references therein).

Estimates of $D_2$ 
from the CMB and the XRB
are consistent with $D_2=3$ to within
$10^{-4}$  on the very
large scales (Peebles 1993; Wu et al. 1999).
While we reject the pure fractal model in this review, the performance
of CDM-like models of fluctuations on large scales have yet to be
tested without assuming homogeneity {\it a priori}. On scales below,
say, $30 \Mpc$, the fractal nature of clustering implies that one has
to exercise caution when using statistical methods which assume
homogeneity (e.g. in deriving cosmological parameters).  
We emphasize that we only considered
one `alternative' here, which is the pure fractal model where $D_2$ is a
constant on all scales.

\section {More Realistic Inhomogeneous Models} 

As the Universe appears clumpy on small scales it is clear that 
assuming the Cosmological Principle and the FRW metric is only an 
approximation, and one has to 
average carefully the density in  
Newtonian Cosmology (Buchert \& Ehlers 1997).
Several models in which the matter in clumpy 
(e.g. 'Swiss cheese' and voids)
have been proposed 
(e.g. Zeldovich 1964; Krasinski 1997; Kantowski 1998; Dyer \& Roeder  
1973; Holz \& Wald 1998;  C\'el\'erier 1999; Tomita 1999). 
For example, if the line-of-sight to a distant  object is `empty' 
it results in a gravitational lensing de-magnification of the object.
This modifies the  FRW luminosity-distance relation, with 
a clumping factor as  another free parameter. 
When applied to a sample of SNIa
the density parameter  of the Universe
$\Omega_{m}$ could be underestimated if FRW is used 
(Kantowski 1998; Perlmutter et al. 1999).
Metcalf \& Silk (1999) pointed out that this effect can be used as a test 
for the nature of the dark matter, i.e. to test if it is smooth  
or clumpy.

\section {A `Best Fit Universe': a Cosmic Harmony ? }

Several groups  (e.g. Eisenstein, Hu \& Tegmark 1998; 
Webster et al. 1998; Gawiser \& Silk 1998;
Bridle et al. 1999) 
have recently estimated cosmological parameters by joint analysis 
of data sets (e.g. CMB, SN, redshift surveys, cluster abundance
and peculiar velocities) in the framework of FRW cosmology.
The idea is that  cosmological parameters can be better estimated
by the complementary of the different probes.

While this approach is promising and we will see more of it in the 
next generation of galaxy and CMB surveys (2dF/SDSS/MAP/Planck)
it is worth emphasizing a  
`health warning' on this approach.
First, 
the choice of parameters space is arbitrary
and in
the Bayesian framework there is freedom in 
choosing a prior for the model.  
Second, the `topology' of the parameter space is only helpful 
when `ridges' of 2 likelihood `mountains' cross each other
(e.g. as in the case of the CMB and the SN). It is more problematic if 
the joint maximum ends up in a `valley'.
Finally, there is the uncertainty that a sample does not represent 
a typical patch of the FRW Universe to yield reliable global cosmological 
parameters.

\subsection{Cosmological Parameters}

Webster et al. (1998) combined results from a 
range of CMB experiments, with a likelihood analysis of the IRAS
1.2Jy survey, performed in spherical harmonics.  
This method expresses the effects of the
underlying mass distribution on both the CMB potential fluctuations
and the IRAS redshift distortion. This breaks the degeneracy 
e.g. between $\Omega_{m}$ and the bias parameter.
The family of CDM models analysed corresponds to a
spatially-flat Universe with an initially scale-invariant
spectrum and a cosmological constant $\lambda$. Free parameters in the joint
model are the mass density due to all matter ($\Omega_{m}$), Hubble's
parameter ($h = H_0 / 100$ km/sec), IRAS light-to-mass bias
($b_{iras}$) and the variance in the mass density field measured in an
$8 h^{-1}$ Mpc radius sphere ($\sigma_{8}$).  For fixed baryon density
$\Omega_b = 0.02/h^2$ the joint optimum lies at 
$\Omega_{m}= 1 - \lambda
= 0.41\pm{0.13}$, $h = 0.52\pm{0.10}$, $\sigma_8 = 0.63\pm{0.15}$,
$b_{iras} = 1.28\pm{0.40}$ (marginalised 1-sigma error bars).
For these values of $\Omega_{m}, \lambda$ and $H_0$
the age of the Universe is $\sim 16.6$ Gyr.

The above parameters correspond to the combination of parameters
$\Omega_{m}^{0.6} \sigma_8 = 0.4 \pm 0.2 $.
This is quite in agreement from results form cluster abundance 
(Eke et al. 1998), 
$\Omega_{m}^{0.5} \sigma_8 = 0.5 \pm 0.1 $.
By combining the abundance of clusters with the CMB and IRAS 
Bridle et al. (1999) found 
$\Omega_{m}= 1 - \lambda
= 0.36$, $h = 0.54$, $\sigma_8 = 0.74$, and
$b_{iras} = 1.08$ (with error bars similar to those  above).

On the other hand, results from peculiar velocities
yield higher values (Zehavi \& Dekel 1999), 
$\Omega_{m}^{0.6} \sigma_8 = 0.8 \pm 0.1$.
By combining the peculiar velocities (from the SFI sample) 
with CMB and 
SN Ia one obtains overlapping likelihoods at  the level of
$2-sigma$ (Bridle et al. 2000). 
The best fit parameters are 
$\Omega_{m}= 1 - \lambda
= 0.42$, $h = 0.63$, and $\sigma_8 = 1.24$.

\subsection{Hyper-Parameters}

A complication that arises in combining data sets 
is that there is freedom
in assigning  the relative weights of different measurements.
A Bayesian approach to the problem utilises  `Hyper Parameters'
(Lahav et al. 2000).

Assume that we have 2 independent data sets, $D_{A}$ and $D_{B}$
(with $N_{A}$ and $N_{B}$ data points respectively) 
and that we wish to determine a vector of free parameters ${\bfw}$
(such as the density parameter $\Omega_{\rm{m}}$, 
the Hubble constant $H_0$ etc.).
This is commonly done by minimising  
\begin{equation}
\chi^2_{\rm joint} = \chi^2_A \; + \; \chi^2_B\; , 
\label{chi2_simple}
\end{equation}
(or maximizing the sum of log  
likelihood functions).
Such procedures assume that the quoted observational random errors 
can be trusted, and that the two (or more) $\chi^2$s  
have equal weights.  
However, when combining `apples and oranges' one may wish to allow freedom in 
the relative weights. 
One possible approach is to generalise Eq. \ref{chi2_simple} to be 
\begin{equation}
\chi^2_{\rm joint} = \alpha  \chi^2_A \; + 
                \beta \; \chi^2_B \; , 
\label{chi2_hp}
\end{equation}
where $\alpha$ and $\beta$ are
`Lagrange multipliers', or `Hyper-Parameters' (hereafter HPs),
which are   to be evaluated 
in a Bayesian way.
There are a number of ways to interpret the meaning of the HPs.
A simple example of the HPs is the case
that 
\begin{equation}
\chi^2_{A} = 
\sum {\frac {1} {\sigma_{i}^{2}}}  [x_{{\rm obs},i} - x_{{\rm pred},i}(\bfw)]^2\;,
\label{chi2_example}
\end{equation}
where the sum is over $N_{A}$ 
measurements and corresponding predictions and errors
$\sigma_{i}$. Hence by multiplying $\chi^2$ by $\alpha$
each  error 
effectively 
becomes ${\alpha}^{-1/2} \sigma_{i}$.
But even if the measurement errors are accurate, 
the HPs are useful in assessing the relative weight of different experiments.
It is not uncommon that astronomers 
discard measurements (i.e. by assigning $\alpha=0$)
in an ad-hoc way. The procedure we propose gives an objective diagnostic 
as to which measurements are problematic 
and deserve further understanding of systematic or random 
errors.   

If the prior probabilities 
for $\ln (\alpha) $ and $\ln (\beta) $ are uniform then 
one should consider the quantity 
\begin{equation}
-2 \; \ln P(\bfw| D_{A}, D_{B}) =   N_{A} \ln 
(\chi_{A}^{2})   + N_{B} \ln 
(\chi_{B}^{2}) 
\label{chi2_new}
\end{equation} 
instead of Eq. \ref{chi2_simple}.

More generally for $M$ data sets one 
should minimise 
\begin{equation}
-2 \; \ln P(\bfw| {\rm data} ) = 
\sum_{j=1}^{M} N_{j} \ln (\chi_{j}^{2}),   
\label{sum_hp}
\end{equation}  
where $N_j$ is the number of measurements in data set $j=1, ..., M$.
It is as easy to calculate this statistic as the standard $\chi^2$.
The corresponding HPs 
can be identified as 
$\alpha_{{\rm eff},j} = N_j/\chi^2_j$ 
(where the $\chi^2_j$'s are evaluated 
at the values of the parameters 
${\bf w}$ that 
minimise eq. \ref{chi2_new})
and they provide useful diagnostics on the reliability of different data sets. 
We emphasize that a low HP assigned to an experiment does not necessarily
mean that the experiment is `bad', but rather it calls attention to look 
for systematic effects or better modelleing. 
The method is illustrated (Lahav et al. 2000) 
by  estimating  the Hubble constant $H_{0}$ from 
different sets of recent CMB experiments (including Saskatoon, Python V, 
MSAM1, TOCO and Boomerang).

\section{Discussion}

Analysis of the CMB, the XRB, radio sources and the Lyman-$\alpha$ 
which probe scales of 
$\sim 100-1000 \Mpc$ strongly support the Cosmological Principle
of homogeneity and isotropy.
They rule out  a pure fractal model.
However, there is a need for more realistic  inhomogeneous models
for the  small scales. This is in particular important for
understanding the validity of cosmological parameters obtained
within the standard FRW cosmology.

Joint analyses of the CMB, IRAS, SN, cluster abundance and
peculiar velocities suggests   $\Omega_{m}=1-\lambda \approx 0.4$.
 With the dramatic increase of data, we should soon be able to map
the fluctuations with scale and epoch, and to analyze jointly LSS 
(2dF, SDSS) and
CMB (MAP, Planck) data, taking into account generalized forms of 
biasing.

\bigskip
{\bf Acknowledgments}
I thank my collaborators for their contribution to the work
 presented here.



\end{document}